\documentclass[aps,pre,twocolumn,superscriptaddress,showpacs]{revtex4}
\usepackage{graphicx}
\usepackage{epsfig, amsmath, amssymb}
\usepackage{color}

\newcommand{\be}{\begin{equation}}
\newcommand{\ee}{\end{equation}}
\newcommand{\beq}{\begin{eqnarray}}
\newcommand{\eeq}{\end{eqnarray}}

\begin{document}
\title{Dynamics of Kuramoto oscillators with time-delayed postitive and negative couplings}
\author{Hui Wu}
\affiliation{Department of Mathematics, Clark Atlanta University, Atlanta, USA}
\author{Mukesh Dhamala}
\affiliation{Department of Physics and Astronomy, Neuroscience Institute,  Georgia State
  and Georgia Tech Center for Advanced Brain Imaging, Georgia State University, Atlanta, USA}
\date{\today}
\pacs{
  05.45.Xt, %Synchronization; coupled oscillators
  02.30.Ks, %Delay and functional equations
  87.10.Ca, %   Analytical theories
 87.19.lj}%Neuronal network dynamics
\begin{abstract}
  Many real-world examples of distributed oscillators involve not only time delays but also attractive (positive) and
  repulsive (negative) influences in their network interactions. Here, considering such examples, we generalize the Kuramoto
  model of globally coupled oscillators with time-delayed positive and negative couplings to explore the effects of such couplings
  in collective phase synchronization. We analytically derive the exact solutions for stable
  incoherent and coherent states in terms of the system parameters allowing us to precisely understand the interplay of time delays and couplings
  in collective synchronization. Dependent on these parameters, fully coherent, incoherent states and mixed
  states are possible. Time-delays especially in the negative coupling seem to facilitate collective synchronization. In case of a stronger
  negative coupling than positive one, a stable synchronized state cannot be achieved without time delays. We discuss the implications
  of the model and the results for natural systems, particularly neuronal network systems in the brain.
\end{abstract}
\maketitle

{\it Introduction.} The Kuramoto model~\cite{Kuramoto:1975}, originally formulated to simplify the Winfree's
biological oscillator model for the circadian rhythms of living systems~\cite{Winfree:1967}, represents an
analytically tractable model to study collective synchronization in large systems of coupled nonlinear oscillators.
Since its conception, the Kuramoto model has been generalized to explain collective dynamical behaviors in many
natural and technological systems of coupled nonlinear oscillators often with various coupling scenarios~
\cite{Pikovsky:2001,Strogatz:2003,Rodrigues:2016,Boccaletti:2016}. As the nature is replete with beautiful complexities
such as self-organized critical phenomena~\cite{Bak:1996} arising from opposing forces, the interplay of positive-negative
time-delayed interactions in the Kuramoto model of oscillators can also be expected to show rich dynamical behaviors yet to be investigated.

The generalizations of the Kuramoto model previously analyzed include coupling scenarios with separate time-delayed interactions,
and positive as well as negative coupling. In an outstanding analytical work, Yeung and Strogatz~\cite{Yeung:1999} considered time delay
in a mean-field sinusoidal coupling of Kuramoto model and derived exact formulas for the stability boundaries of the
incoherent and synchronized states as a functional of the delay in a special case where the oscillators were identical.
Later, Earl and Strogatz~\cite{Earl:2003} extended the analysis to a variety of coupling topologies (regular, small-world and random network graphs)
to find that the same stability criterion held true as for the mean-field case. In another more recent study, Hong and Strogatz~\cite{Hong:2011,Hong:2011b}
studied the Kuramoto model with positive and negative coupling parameters (without time-delay) and reported a variety of dynamical
behaviors including fully synchronized, partially synchronized, desynchronized and traveling states. Qui and colleagues~\cite{Qui:2016} recently
discovered a new non-stationary state (named {\it Bellerophone}) along with a synchronized state in the model with positive and negative coupling.
However, the dynamics of the Kuramoto model with combined time-delayed interactions and positive-negative (attractive-replusive or excitatory-inhibitory)
couplings have largely remained unexplored despite their relevance to many natural systems, such as biological, physical, chemical and social networks.

Interaction time-delays and excitation-inhibition (E-I) are usually characteristics of spatially distributed, self-organized
systems, such as neurons in the brain in which the E-I balance is required to maintain normal temporal and
spatial functional organization in healthy cognitive functions and behaviors~\cite{WilsonCowen:1972,Wang:1996,Brunel:2003,Buzsaki:2006,Mann:2010}.
Networks of coupled excitatory and inhibitory neurons can exhibit a complex dynamical behaviors including synchronization, multiclustered solutions,
oscillator death~\cite{Stefan:2008}. E-I couplings in Belousov-Zhabotinsky can lead to a large number of spatiotemporal patterns,
simple to complex behaviors~\cite{Vanag:2011}. In social networks of conformists (dynamical units with positive coupling) and contrarians
(those with negative coupling) also, various collective behaviors are possible~\cite{Galam:2004,Lama:2005,Hong:2011b}.
Interactions in spatially distributed oscillator network systems are not generally instantaneous. On the contrary, finite speed of signal
transmission over a distance gives rise to a finite time delay. For example, signal transmission time delays are inherent in networks of
neurons in the brain. While the chemical synaptic time delays are small ($\sim$2 ms), the axonal conduction delays, which depend on the distance
between neurons in the brain, can reach upto tens of milliseconds~\cite{Izhikevich:2006}. Time delays comparable to time-scales of neuronal
oscillations are known to have significant effects in the collective (ensemble) activity of neurons~\cite{Dhamala:2004,Izhikevich:2006,Adhikari:2011}.
Time delays in networks of coupled inhibitory neurons can induce a variety of phase-coherent dynamic behaviors~\cite{Liang:2009}. Time delays in
networks with E-I can be expected to induce a variety of collective dynamical behaviors.

In this work, we analyze a generalized Kuramoto model of nonlinear oscillators with time-delayed positive and negative couplings for
stability of coherent states and incoherent states and derive the exact analytical solutions to define the parameter boundaries for these states.
As general results, we come to show that fully coherent, incoherent states and mixed states are possible, and time delays facilitate
synchronized states in case of a stronger combined negative coupling than positive one.

{\it Methods and results}. We start with the following generalized Kuramoto model of $N$ oscillators globally connected with time-delayed
positive and negative couplings:

\begin{eqnarray}\nonumber\label{eq:0}
\dot{\theta_i}(t)= w_0+\frac{k_1}{N_1}\sum_{j=1}^{N_1}f(\theta_j(t-\tau_1)-\theta_i(t))\\ +\frac{k_2}{N_2}\sum_{j=N_1+1}^{N}f(\theta_j(t-\tau_2)-\theta_i(t))
\end{eqnarray}
Here, the coupled system consists of $N_1$ and $N_2$ numbers of positively and negatively coupled oscillators. The total number is $N = N_1+N_2$, and
$k_1+k_2=(1-c)k:=K$. $k$ is the coupling strength, $k_1$ is a positive coupling strength, $k_2$ is a negative coupling strength, and $c=k_2/k$ is the coupling ratio.
$\theta_i(t)$ is the phase of the ith oscillator, $\dot{\theta_i}(t)$ its derivative and each oscillator has the same natural frequency $w_0$. $f$ is a
general coupling function, which we replace with a sinusoidal function in our examples below. $\tau_1$ is the time delay associated with positive coupling and
$\tau_2$ is the time delay with negative coupling. Below, we explore the interplay of the parameters ($k$, $c$, $\tau_1$ and $\tau_2$) in the emergence of
stable coherent states.

The rest of the paper is organized as follows: (a) we analyzed the stability of a completely coherent state and derive the boundary curves for the coherent state, (b) we
analyzed the stability of a completely incoherent state and derive the boundary curves for the in coherent state. In each case, we consider the following
scenarios of delays (i)$\tau_1=\tau_2=\tau$, and (ii)$\tau_1=\tau>0$ and $\tau_2=0$. For the first case (a), we analyzed the time-evolution of a
perturbation to a synchronized state. For the second case (b), we linearize the continuity equation around the incoherent state ($\rho(\theta,\omega,t)=1/2\pi$)
describing the time-evolution of instantaneous phase distribution $\rho(\theta,\omega,t)$ on a unit circle, analyzed the effect of a perturbation on
the incoherent state.

{\it (a) Stability Analysis of coherent (synchronized) state}. We assume the system reaches a complete synchronized state described by the following equation:
\begin{eqnarray}\label{eq:1}
\theta_i(t)=&\Omega t
\end{eqnarray}
The collective frequency $\Omega$ is given by
\begin{eqnarray}\label{eq:1.1}
\Omega&=&w_0+k_1f(-\Omega\tau_1)+k_2f(-\Omega\tau_2)
\end{eqnarray}
Now, we add a small perturbation term on (\ref{eq:1}):

\begin{eqnarray}\label{eq:2}
\theta_i(t)=&\Omega t+\epsilon\phi_i(t),
\end{eqnarray}
where $0<\epsilon\ll 1$. Now, if we define $\Delta_i(t):=\phi_i(t)-\phi_1(t)$:
\begin{eqnarray}\nonumber\label{eq:5}
  \dot{\Delta}_i(t)&=& -\frac{k_1}{N_1}f'(-\Omega\tau_1)\sum_{j=1}^{N_1}\Delta_i(t)-\frac{k_2}{N_2}f'(-\Omega\tau_2)\sum_{j=N_1+1}^{N}\Delta_i(t)\\
  &=&-[k_1f'(-\Omega\tau_1)+k_2f'(-\Omega\tau_2)]\Delta_i(t)
\end{eqnarray}

The stability requirement for synchronization is that $\Delta_i(t)$ be decaying to zero instead of diverging to infinity. Hence,
\begin{eqnarray}\label{eq:6}
k_1f'(-\Omega\tau_1)+k_2f'(-\Omega\tau_2)>0
\end{eqnarray}

When (\ref{eq:6}) is satisfied for $\forall i=1,2,3,...N$, $\phi_i(t)$ exponentially converges to the same function $\phi(t)$. Thus, $\phi(t)$ satisfies the
following time delay equation:
\begin{eqnarray}\nonumber\label{eq:7}
\dot{\phi}(t)= \frac{k_1}{N_1}f'(-\Omega\tau_1)\sum_{j=1}^{N_1}[\phi(t-\tau_1)-\phi(t)]\\ +\frac{k_2}{N_2}f'(-\Omega\tau_2)\sum_{j=N_1+1}^{N}[\phi(t-\tau_2)-\phi(t)]
\end{eqnarray}

If we let $\tau_1=\tau_2=\tau$ for a synchronization state, we have
\begin{eqnarray}\label{eq:8}
(1-c)f'(-\Omega\tau)>0
\end{eqnarray}
\begin{eqnarray}\nonumber\label{eq:9}
\dot{\phi}(t)=(1-c)kf'(-\Omega\tau)[\phi(t-\tau)-\phi(t)]
\end{eqnarray}

With $\phi(t)=e^{\lambda t}$ to find the characteristic function, we have
\begin{eqnarray}\label{eq:10}
\lambda=(1-c)kf'(-\Omega\tau)(e^{-\lambda \tau}-1)
\end{eqnarray}

A brief analysis shows that when (\ref{eq:8}) is satisfed, for this type of characteristic function and $\lambda\neq 0$, all eigenvalues have
negative real parts, i. e. $Re(\lambda)< 0$ (stable state).

If we let $\tau_1=\tau>0$, $\tau_2=0$, (\ref{eq:6}) becomes
\begin{eqnarray}\label{eq:12}
f'(-\Omega\tau)-cf'(0)>0
\end{eqnarray}

We now let the general coupling function to be a sinusoidal, $f(\theta)=\sin(\theta)$, for $\tau_1=\tau_2=\tau$ and we arrive at the following two cases:
(i) if $1-c>0$, then $\cos(\Omega\tau)>0$, and (ii) if $1-c<0$, then $\cos(\Omega\tau)<0$.
%%%%%%%%%%%%%%%%%%%%%%%%%%%%%%%%%%%%%%%%%%%%%
%\begin{figure}
%\epsfig{figure=Fig1.eps,width=1.0\linewidth}
%\caption{(Color online). Stable and unstable points.}
%\label{fig:fig1}
%\end{figure}
%%%%%%%%%%%%%%%%%%%%%%%%%%%%%%%%%%%%%%%%%%%%

We let $T=\frac{2\pi}{\omega_0}$, $y=\frac{k}{\omega_0}$ and arrive at the solutions for the boundary curves of the unstable coherent states in two cases:
(i) when $1-c>0$, we have
\begin{eqnarray}\label{eq:20}
\frac{4n+1}{4(1-(1-c)y)}<\frac{\tau}{T}<\frac{4n+3}{4(1+(1-c)y)}
\end{eqnarray}
and $0<y<\frac{1}{2(1-c)(2n+1)}$.

and (ii) when $1-c<0$, we have

\begin{eqnarray}\label{eq:21}
\frac{4n-1}{4(1+(1-c)y)}<\frac{\tau}{T}<\frac{4n+1}{4(1-(1-c)y)}
\end{eqnarray}

and $0<y<\frac{1}{4n(c-1)}$. We use $(\ref{eq:20})$ and $(\ref{eq:21})$ to draw the boundary curves and areas for unstable
synchronized states, as shown in black shades of Fig. 1(a).

We now consider $f(\theta)=\sin(\theta)$, $\tau_1=\tau>0$,  $\tau_2=0$. According to $(\ref{eq:12})$,
\begin{eqnarray}\label{eq:22}
\cos(\Omega\tau)-c>0
\end{eqnarray}
is required to have a stable solution. This solution $(\ref{eq:12})$ implies that a time delay is needed for a stable synchronization in case of
a stronger negative coupling ($c>1$).  
When $\cos(\Omega\tau)-c>0$, $\forall i=1,2,...N$: $\phi_i(t)$ exponentially
converges to the same $\phi(t)$, $\phi(t)$ then satisfies the following equation:

\begin{eqnarray}\nonumber\label{eq:23}
\dot{\phi}(t)= k\cos(\Omega\tau)[\phi(t-\tau)-\phi(t)]
\end{eqnarray}

Then, the characteristic function satisfies:
\begin{eqnarray}\label{eq:24}
\lambda=k\cos(\Omega\tau)(e^{-\lambda \tau}-1)
\end{eqnarray}

It is easy to show that for any $k>0$ and $\cos(\Omega\tau)>c$, for all $\lambda\neq 0$, $Re(\lambda)<0$. Therefore, the
synchronization state is always linearly stable.
Whenever $c\ge 1$, we do not have a stable coherent state. Taking $y=\frac{k}{\omega_0}$, $T=\frac{2\pi}{\omega_0}$,
when $0<c<1$, there is no stable
synchronization state if and only if $\tau$ satisfies following inequality:
\begin{eqnarray}\label{eq:26}
\frac{2n\pi+arc\cos c}{2\pi(1-\sqrt{1-c^2}y)}<\frac{\tau}{T}
< \frac{2(n+1)\pi-arc\cos c}{2\pi(1+\sqrt{1-c^2}y)}
\end{eqnarray}
and
\begin{eqnarray}\label{eq:27}
0<y<\frac{\pi-arc\cos c}{(2n+1)\pi\sqrt{1-c^2}}
\end{eqnarray}
We use $(\ref{eq:24})$ and $(\ref{eq:26})$ to draw the boundary curves and areas for this unstable coherent state, as shown with black shade in Fig. 1(b).
%%%%%%%%%%%%%%%%%%%%%%%%%%%%%%%%%%%%%%%%%%%%
\begin{figure}
\epsfig{figure=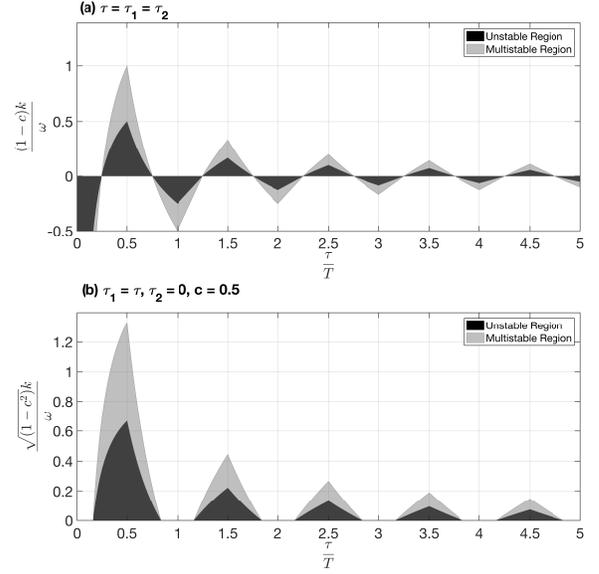,width=1\linewidth}
\caption{Regions of unstable coherent and stable incoherent states (dark shade), stable coherent and incoherent states (gray), and stable
  coherent state (unshaded or white) as a function of coupling strength and time delay. These boundaries of stability obtained by
  the analytical solutions were also numerically verified.}
\label{fig:fig2}
\end{figure}
%%%%%%%%%%%%%%%%%%%%%%%%%%%%%%%%%%%%

{\it (b) Stability analysis of incoherent state}.

We now approach the stability of incoherent state, from equation (\ref{eq:0}), by considering the mean field of positive oscillators
and the mean field of negative oscillators:
\begin{eqnarray}\nonumber\label{eq:28}
r_1e^{i\phi_1}&=&\frac{1}{N_1}\sum_{j=1}^{N_1}e^{i\theta_j}\\
r_2e^{i\phi_2}&=&\frac{1}{N_2}\sum_{j=N_1+1}^{N_1+N_2=N}e^{i\theta_j}
\end{eqnarray}

The conditional probability density function $\rho(\theta, \omega, t)$ satisfies
\begin{eqnarray}\label{eq:29}
\int_0^{2\pi}\rho(\theta,\omega,t)d\theta=1
\end{eqnarray}

For complete incoherent state:
\begin{eqnarray}\label{eq:29:1}
\rho(\theta,\omega,t)=\frac{1}{2\pi}
\end{eqnarray}

For $\forall$ $i=1,2,3,...N$,
\begin{eqnarray}\label{eq:30}
r_1e^{i(\phi_1(t-\tau_1)-\theta_i(t))}=\frac{1}{N_1}\sum_{j=1}^{N_1} e^{i(\theta_j(t-\tau_1)-\theta_i(t))}
\end{eqnarray}

Similarly,
\begin{eqnarray}\label{eq:31}
r_2e^{i(\phi_2(t-\tau_2)-\theta_i(t))}=\frac{1}{N_2}\sum_{j=N_1+1}^{N} e^{i(\theta_j(t-\tau_2)-\theta_i(t))}
\end{eqnarray}

Considering the imaginary part of (\ref{eq:30}) and (\ref{eq:31}), we can transfer the dynamical equation in the mean field form as:

\begin{eqnarray}\nonumber\label{eq:33}
\dot{\theta_i}(t)= \omega_i+k_1r_1\sin(\theta_j(t-\tau_1)-\theta_i(t))\\
+k_2r_2\sin(\theta_j(t-\tau_2)-\theta_i(t))
\end{eqnarray}

The equation with a perturbation to the incoherent state is then:
\begin{eqnarray}\label{eq:34}
\rho(\theta,\omega,t)=\frac{1}{2\pi}+\epsilon \eta(\theta,\omega,t)
\end{eqnarray}
with $\epsilon\ll 1$.

Here $\eta(\theta,\omega,\tau)$ can be expanded into following Fourier series:
\begin{eqnarray}\label{eq:42}
\eta(\theta,\omega,t)=c(\omega,t)e^{i\theta}+c^*(\omega,t)e^{-i\theta}+\eta^{\perp}(\theta,\omega,t)
\end{eqnarray}

Here, $\eta^{\perp}$ are higher Fourier Harmonics. We now look for a type of solution of the form:
\begin{eqnarray}\label{eq:47}
c(\omega,t)=b(\omega)e^{\lambda t}
\end{eqnarray}
\begin{eqnarray}\nonumber\label{eq:48}
\lambda b(\omega)=-i\omega b(\omega)+\frac{k_1e^{-\lambda\tau_1}}{2}\int_{-\infty}^{\infty}b(v)g(v)dv\\+\frac{k_2e^{-\lambda\tau_2}}{2}\int_{-\infty}^{\infty}b(v)g(v)dv
\end{eqnarray}

For identical frequencies,
$\omega_i=\omega_0$, $\forall i=1,2,...N$, $g(\omega)=\delta(\omega-\omega_0)$, $\delta$ is the dirac delta function.

The eigenvalue $\lambda$ satisfies the following equation:
\begin{eqnarray}\label{eq:49}
\lambda +i\omega_0=\frac{1}{2}(k_1e^{-\lambda\tau_1}+k_2e^{-\lambda\tau_2})
\end{eqnarray}

When $\tau_1=\tau_2=\tau$, $k_1=k>0$, $k_2=-ck<0$, $c>0$ is a constant,
\begin{eqnarray}\label{eq:50}
(1-c)k=2(\lambda+i\omega_0)e^{\lambda\tau}
\end{eqnarray}

At a critical (bifurcation) point $K_c$, the eigenvalue $\lambda$ passes through the imaginary axis $\lambda=R i$, $R$ is real number.\

From the equation above,
$2(\lambda+i\omega_0)e^{\lambda\tau}=2i(R+\omega_0)(\cos(R\tau)+i\sin(R\tau))$ has to be a real value.
Hence, $\cos(R\tau)=0$. If we take $R\tau=(-m+\frac{1}{2})\pi$, $m\in Z$.
\begin{eqnarray}\label{eq:51}
(1-c)k=2(-1)^{m-1}(R+\omega_0)
\end{eqnarray}
\begin{eqnarray}\nonumber\label{eq:52}
2\omega_0\tau+(-1)^m(1-c)k\tau=-2R\tau\\=2(m-\frac{1}{2})\pi=(2m-1)\pi
\end{eqnarray}

Taking $m=2n+1$ odd positive integers, we get one boundary for $\tau$:

\begin{eqnarray}\label{eq:54}
\tau_{c1}=\frac{(4n+1)\pi}{2\omega_0-(1-c)k}
\end{eqnarray}

If we take $m=2n$ or $m=2(n+1)$ even positive integers, we get other boundaries for $\tau$:

\begin{eqnarray}\label{eq:55}
\tau_{c2}=\frac{(4n-1)\pi}{2\omega_0+(1-c)k}
\end{eqnarray}

\begin{eqnarray}\label{eq:55-1}
\tau_{c3}=\frac{(4n+3)\pi}{2\omega_0+(1-c)k}
\end{eqnarray}

Letting $T=\frac{2\pi}{\omega_0}$, $y=\frac{k}{w_0}$, for $(1-c)k=K>0$ and $\tau_{c1}<\tau<\tau_{c3}$, we have the stable
incoherent state within:
\begin{eqnarray}\label{eq:56-01}
\frac{4n+1}{4(1-\frac{(1-c)y}{2})}<\frac{\tau}{T}<\frac{4n+3}{4(1+\frac{(1-c)y}{2})}
\end{eqnarray}

For $(1-c)k=K<0$, $\tau_{c2}<\tau<\tau_{c1}$, we have the stable incoherent state within:
\begin{eqnarray}\label{eq:56-02}
\frac{4n-1}{4(1+\frac{(1-c)y}{2})}<\frac{\tau}{T}<\frac{4n+1}{4(1-\frac{(1-c)y}{2})}
\end{eqnarray}

We use $(\ref{eq:56-01})$ and $(\ref{eq:56-02})$ to draw the boundary curves and areas for this unstable incoherent state, as shown with grey shade in Fig. 1(a).

For $\tau_1=\tau>0$, $\tau_2=0$, the eigenvalue $\lambda$ has to satisfy the following equation:
\begin{eqnarray}\label{eq:57}
\lambda +i\omega_0=\frac{1}{2}(k_1e^{-\lambda\tau}+k_2)
\end{eqnarray}

If $c\ge 1$, let $\lambda=\alpha+iR$, $\alpha, R$ are real.

\begin{eqnarray}\label{eq:59}
k(e^{-\alpha\tau}\cos(R\tau)-c)=2\alpha
\end{eqnarray}

It is easy to see that when $c>1$, $(\ref{eq:59})$ requires $\alpha<0$, while $c=1$ and $\alpha\ge 0$, then the only possibility
to satisfy $(\ref{eq:59})$ is $\alpha=R=0$. So, the incoherent state is always neutrally stable whenever $c\ge 1$.
An unstable incoherent state only possibly exist for $0<c<1$.

At a bifurcation point, $\lambda=Ri$, R is real. Substituting it to $(\ref{eq:57})$:
\begin{eqnarray}\label{eq:60}
2(R+\omega_0)i=k(\cos(\tau R)-i\sin(\tau R))-ck
\end{eqnarray}
\begin{eqnarray}\label{eq:61}
(2R+2\omega_0+k\sin(\tau R))i=k(\cos(\tau R)-c)
\end{eqnarray}

The above equation is satified if and only if $\cos(\tau R)=c$ and
\begin{eqnarray}\label{eq:62}
k\sin(\tau R)=-2(R+\omega_0)
\end{eqnarray}
$\tau R=-2m\pi\pm arc\cos c$, $m$ are integers. $\sin(\tau R)=\pm\sqrt{1-c^2}$.

\begin{eqnarray}\label{eq:63}
k=\mp\frac{2(R+\omega_0)}{\sqrt{1-c^2}}
\end{eqnarray}

If we take $m=n$ and $\tau R=-2n\pi- arc\cos c$, we get the left boundary for $\omega_0\tau$ as:

\begin{eqnarray}\label{eq:64}
\omega_0\tau_{lc}=2n\pi+arc\cos c+\frac{1}{2}k\tau\sqrt{1-c^2}
\end{eqnarray}

Similarly, if take $m=n+1$ and $\tau R=-2n\pi+ arc\cos c$, we get the right boundary for $\omega_0\tau$ as:
\begin{eqnarray}\label{eq:65}
\omega_0\tau_{rc}=2(n+1)\pi-arc\cos c-\frac{1}{2}k\tau\sqrt{1-c^2}
\end{eqnarray}

The incoherent state is stable if and only if $\omega_0\tau_{lc}<\omega_0\tau<\omega_0\tau_{rc}$. If make $y=\frac{k}{\omega_0}$, we have
\begin{eqnarray}\label{eq:67}
\frac{2n\pi+arc\cos c}{2\pi(1-\frac{\sqrt{1-c^2}y}{2})}<\frac{\tau}{T}<\frac{2(n+1)\pi-arc\cos c}{2\pi(1+\frac{\sqrt{1-c^2}y}{2})}
\end{eqnarray}

We use $(\ref{eq:65})$ and $(\ref{eq:67})$ to draw the boundary curves and areas for this unstable incoherent state, as shown with grey shade in Fig. 1(b).

Finally, we have also numerically verified all these analytical results of Fig.1(a,b).

{\it Conclusions and discussion}. Here, we generalize the Kuramoto model of globally coupled phase oscillators
with time-delayed positive-negative coupling. We have analytically and numerically studied the stability of synchronized and incoherent states
in this generalized Kuramoto model. We derive the exact solutions for the critical coupling strengths at different
time delays for stable incoherent and coherent states. These derivations proivde new insights into how the interplay of time delays
can affect collective synchronization. We find that fully coherent, incoherent and mixed states are possible dependent on the parameters
in this generalized model. Time delayed interactions are helpful for achieving synchronization in the case of dominant negative network coupling.
We expect that our theoretical work will be useulf for further research in trying to understand the role of excitation-inhibition in self-organized systems
of distributed oscillators like the neuronal systems in the brain.


\begin{references}
\bibitem{Kuramoto:1975}
Y. Kuramoto, in {\it Proceedings of the International Symposium
on Mathematical Problems in Theoretical Physics}, edited by H. Araki, Lecture Notes in Physics Vol. 39
(Springer, Berlin, 1975); {\it Chemical Oscillations, Waves, and Turbulence} (Springer, Berlin, 1984).
\bibitem{Winfree:1967}
A. T. Winfree, J. Theor. Biol. {\bf 16}, 15 (1967).
\bibitem{Pikovsky:2001}
A. Pikovsky, M. Rosenblum, J. Kurths, in {\it Synchronization: a Universal Concept in Nonlinear Sciences}, Cambridge University Press, Cambridge, England, 2001).
\bibitem{Strogatz:2003}
S. Strogatz, {\it Sync: The emerging science of spontaneous order} (Hyperion, New York, 2003).
\bibitem{Rodrigues:2016}
F. A. Rodrigues, T. K. DM. Peron, P. Ji, and J. Kurths, Phys. Rep. {\bf 610}, 1 (2016).
\bibitem{Boccaletti:2016}
S. Boccaletti, J. A. Almendral, S. Guan, I. Leyva, Z. Liu, I. Sendina-Nadal, Z. Wang, and Y. Zou, Phys. Rep. {\bf 660}, 1 (2016).
\bibitem{Bak:1996}
P. Bak, {\it How nature works: the science of self-organized criticality} (Springer-Verlag, New York, 1996).
\bibitem{Yeung:1999}
M. K. Stephen Yeung, S. H. Strogatz, Phys. Rev. Lett. {\bf 82}, 648 (1999).
\bibitem{Earl:2003}
M. G. Earl, S. H. Strogatz, Phys. Rev. E {\bf 67}, 036204-1 (2003).
\bibitem{Hong:2011}
  H. Hong, S. H. Strogatz, Phys. Rev. Lett. {\bf 106}, 054102 (2011).
\bibitem{Hong:2011b}
  H. Hong, S. H. Strogatz, Phys. Rev. E {\bf 84}, 046202 (2011).
\bibitem{Qui:2016}
T. Qui, S. Boccaletti, I. Bonamassa, Y. Zou, J. Zhou, Z. Liu, S. Guan, Sci. Rep. {\bf 6}, 36713 (2016).
\bibitem{WilsonCowen:1972}
H. R. Wilson, J. D. Cowan, Biophys. J. {\bf 12}, 1 (1972).
\bibitem{Wang:1996}
X. J. Wang, G. Buzsáki, J. Neurosci. {\bf 16}, 6402 (1996).
\bibitem{Brunel:2003}
N. Brunel, X. J. Wang, J. Neurophysiol. 90, 415–430 (2003).
\bibitem{Buzsaki:2006}
 G. Buzsaki, in {\it Rhythms of the Brain}, (Oxford University Press, Cambridge, England, 2006).
\bibitem{Mann:2010}
  E. O. Mann, I. Mody, Nat. Neurosci. {\bf 13}, 205 (2010).
\bibitem{Vanag:2011}
V. K. Vanag, I. R. Epstein, Phys. Rev. E {\bf 84} 066209 (2011).
 %"Excitatory and inhibitory coupling in a one-dimensional array of Belousov-Zhabotinsky micro-oscillators: Theory."
\bibitem{Stefan:2008}
  R. A. Stefanaescu, V. K. Jirsa, PLOS Comp. Biol. {\bf 4}, e1000219 (2008).
\bibitem{Galam:2004}
  S. Galam, Physica A {\bf 333}, 453 (2004).
\bibitem{Lama:2005}
  M. S. Lama, J. M. Lopez, H. S. Wio, Europhys. Lett. {\bf 72}, 851 (2005).
\bibitem{Izhikevich:2006}
  E. M. Izhikevich, Neural Comput {\bf 18}, 245-282 (2006).
\bibitem{Adhikari:2011}
  B. M. Adhikari, A. Prasad, M. Dhamala, Chaos {\bf 21}, 023116 (2011).
\bibitem{Dhamala:2004}
  M. Dhamala, V. K. Jirsa, M. Ding, Phys Rev Lett {\bf 92}, 074104 (2004).
\bibitem{Liang:2009}
X. Liang, M. Tang, M. Dhamala, and Z. Liu, Phys. Rev. E {\bf 80}, 066202 (2009).

\end{references}
\end{document}